\documentclass[usenatbib]{mn2e}
\usepackage{epsfig}
\usepackage{amsmath,amssymb,mathrsfs}
\usepackage{aas_macros}

\usepackage[dvips]{color}
\definecolor{rev}{rgb}{0.8,0.0,0.0}
\definecolor{ans}{rgb}{0.0,0.0,1.0}
\definecolor{cut}{rgb}{1.0,0.547,0.0}

\newcommand{\percc}{{\rm cm^{-3}}}

\newcommand{\E}[1]{\times 10^{#1}}

\newcommand{\Rcr}{{\cal R}_{\rm cr}}

\newcommand{\nH}{n_{{\rm H}}}

\newcommand{\Msun}{{\rm M_{\odot}}}

\title[Voronoi particle splitting]
{Particle splitting in smoothed particle hydrodynamics 
based on Voronoi diagram}

\author[G. Chiaki \& N. Yoshida]
{Gen Chiaki$^{1}$\thanks{E-mail: gen.chiaki@utap.phys.s.u-tokyo.ac.jp} and
Naoki Yoshida$^{1,2}$
\\
$^{1}$Department of Physics, Graduate School of Science, The University of Tokyo, 
7-3-1 Hongo, Bunkyo, Tokyo 113-0033, Japan \\
$^{2}$Kavli Institute for the Physics and Mathematics of the Universe (WPI), 
Institutes for Advanced Study, \\
The University of Tokyo, Kashiwa, Chiba 277-8583, Japan}
\begin{document}

\date{}

\pagerange{\pageref{firstpage}--\pageref{lastpage}} \pubyear{2013}

\maketitle

\label{firstpage}

\begin{abstract}
We present a novel method for particle splitting in smoothed particle 
hydrodynamics simulations.
Our method utilizes the Voronoi diagram for a given particle set 
to determine the position of fine daughter particles.  
We perform several test simulations to compare our method 
with a conventional splitting method
in which the daughter particles are placed isotropically 
over the local smoothing length.
We show that, with our method, the density deviation after splitting 
is reduced by a factor of about two compared with the
conventional method. Splitting would smooth out
the anisotropic density structure if the daughters are distributed
isotropically, 
but our scheme allows the daughter particles to
trace the original density distribution with length scales of the 
mean separation of their parent. 
We apply the particle splitting to simulations of the
primordial gas cloud collapse. The thermal evolution is accurately
followed to the hydrogen number density of $10^{12} \ {\rm cm}^{-3}$.
With the effective mass resolution of $\sim 10^{-4} \ \Msun$ after the
multi-step particle splitting, the protostellar disk structure is well resolved.
We conclude that the method offers an efficient way to simulate
the evolution of an interstellar gas and the formation of stars.
\end{abstract}

\begin{keywords} 
gravitation --- 
hydrodynamics ---
methods: numerical ---
stars: formation --- 
stars: Population III
\end{keywords}


\section{INTRODUCTION}
\label{sec:intro}

Smoothed particle hydrodynamics (SPH) is a widely used technique 
to follow the dynamics of a gravitationally interacting gas.
A notable advantage of SPH is that 
the resolution automatically increases
with SPH particles concentrating in dense regions.
Nevertheless, in simulations of some hydrodynamical problems such as 
prestellar gas cloud collapse, one typically
needs to follow a density evolution over too many orders of magnitude.
In the central densest region, SPH particles would eventually violate a resolution requirement.
\citet{Truelove97} argue that a local Jeans length 
needs to be resolved by several grid cells at any time in a grid based
hydrodynamics simulation.
For a SPH simulation with the number of the neighbor particles 
$N_{\rm ngb}$ and
the particle mass $m_{\rm p}$, the resolution criterion may be written as
$M_{\min} < M_{\rm Jeans} / \Rcr$, where
$M_{\min} = N_{\rm ngb} m_{\rm p}$ is the minimal mass resolved 
by the SPH particles, and
$M_{\rm Jeans} \sim c_s^3 G^{-3/2} \rho ^{-1/2}$ is the Jeans mass
at the region with the sound speed $c_s$ and the density $\rho$.
Given a critical mass ratio $\Rcr$, one can determine the maximal particle mass required to resolve 
each simulated region.
Insufficient mass resolution results in producing
artificial features such as smoothed round shape, 
and often triggers unphysical fragmentation 
\citep{Truelove97, Truelove98}. 
Besides, it is costly
to perform simulations with a uniform, and
sufficiently high mass 
resolution so that the entire simulation region is 
fully resolved throughout the run.

\begin{figure}
\includegraphics[width=8cm]{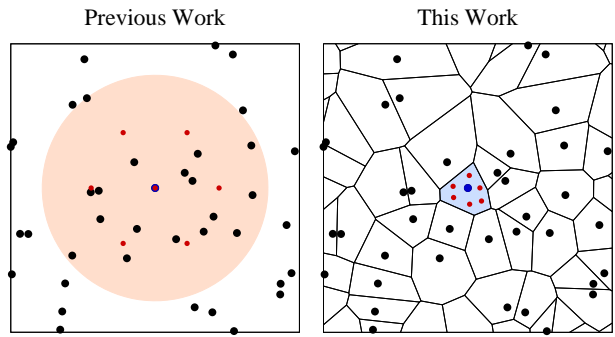}
\caption{
We compare the isotropic 
particle splitting scheme (left) and 
our new scheme based on the Voronoi diagram (right) 
in two dimension.
In the left panel, the central particle (blue point) is 
split to seven daughter particles
(red points) distributed within the smoothing length of 
the parent particle (orange-shaded region).
In the right panel, the daughter particles are placed 
within the Voronoi cell 
of the central particle (blue-shaded region).}
\label{fig:demo2D}
\end{figure}

To save the computational time and memory, multi-step 
zoom-in techniques have been developed.
Mesh refinement can be done in a conceptually simple manner
in grid-based simulations,  
where the cell division procedure can be
systematically defined.
In adaptive mesh refinement (AMR) 
simulations, for example, an initially large
cell can be progressively refined with just 
smaller size rectangular cells.
On the other hand, refinement of SPH particles is not trivial.
A promising refinement technique for SPH simulations 
is particle splitting.
A coarse parent particle which is about to violate 
certain resolution criteria is replaced
with a number of finer daughter particles.
A simple way of particle splitting is to distribute
daughter particles within the smoothing
length $h_{\rm parent}$ of the parent particle
\citep[e.g.][]{Kitsionas02, Bromm03}.
The splitting method of \citet{Kitsionas02} 
(hereafter, KW02) is widely used in the simulations
of the super-massive blackhole formation \citep{Dotti07, Mayer10} and
of primordial star formation \citep{Yoshida06, Hirano14}.
The KW02 method distributes thirteen daughters on a hexagonal closest 
packing array at a distance 
$l_{\rm daughter} = 1.5 \ h_{\rm parent}/13^{1/3}$ from the central particle.
One of the daughters is located the same coordinate as the parent, and
the angular orientation of the twelve daughters is randomly determined.
Since there are, by definition, $N_{\rm ngb}$ particles within $h_{\rm parent}$,
it can happen that a daughter particle
is placed very close to one of the neighboring 
particles by chance. This can result in 
significant overestimation of the local density.
Also the fine anisotropic structures with sizes less 
than $\sim h_{\rm parent}$ tend to be smoothed 
out by imposing the spherical distribution on the daughter particles.
Figure \ref{fig:demo2D} illustrates an example particle distribution. 
The left panel 
shows the two-dimensional 
coordinates of the daughters generated
by the central coarse particle.
After splitting with a random angle orientation, 
several daughters (red dots) happen to be close to 
neighbors (black dots) within the smoothing
kernel (orange-shaded region).
In addition, daughters distributed in the originally 
underdense region can spuriously boost the local density estimate.

The density perturbations would be mitigated
or would cause only local effects if
the daughter particles are placed in the region dominated 
essentially by a single parent particle.
Along this idea, \citet{Martel06} (MES06) present a method in which 
the daughter particles
are placed on the vertices of the cube centered at a target parent particle
with the cube side-length being the half inter-particle distance.
We here explore yet another, and an elegant 
space partitioning method based on 
the Voronoi tessellation.
A Voronoi cell is defined as a set of the points 
closest to each particle.
The right panel of Figure \ref{fig:demo2D} illustrates 
our scheme of particle splitting in two dimension.
The daughter particles (red dots) are distributed within 
a Voronoi cell (defined by the black lines).
They are expected to reproduce the 
original density structure of the coarse parent
particles. A nice feature of the method is that
the refinement can be done in a systematic manner 
for the given distribution of the original coarse particles.
A number of advantages of using the Voronoi tessellation 
have been shown in previous studies.
For example, a moving mesh code {\sc arepo} \citep{Springel10}
and a mesh-free code {\sc gizmo} \citep{Hopkins14} implement 
the particle refinement based on the Voronoi tessellation.

In the present paper, we first apply the particle splitting 
based on the Voronoi diagrams to SPH simulations.
We examine how our method improves the effective resolution
while preserving original density structures, 
compared with a conventional splitting method of KW02 
by several tests.
As a realistic application, we perform simulations of gravitational
collapse of a primordial gas cloud with two splitting methods.

\section{Splitting Method}

\begin{figure}
\includegraphics[width=8cm]{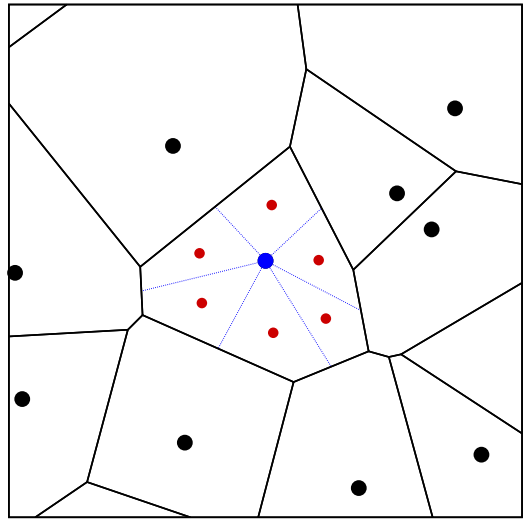}
\caption{
Two-dimensional illustration of the coordinates of daughter particles 
(red dots) around the parent particle
(blue dot). 
A daughter is on the center of mass of a subcell
defined by a blue-dotted line which connects the parent and
the middle point of each Voronoi edge.
The plot is a zoom-in to the central portion
of the right panel in Figure 1.
}
\label{fig:split_demo_2D}
\end{figure}

\begin{figure}
\includegraphics[width=8cm]{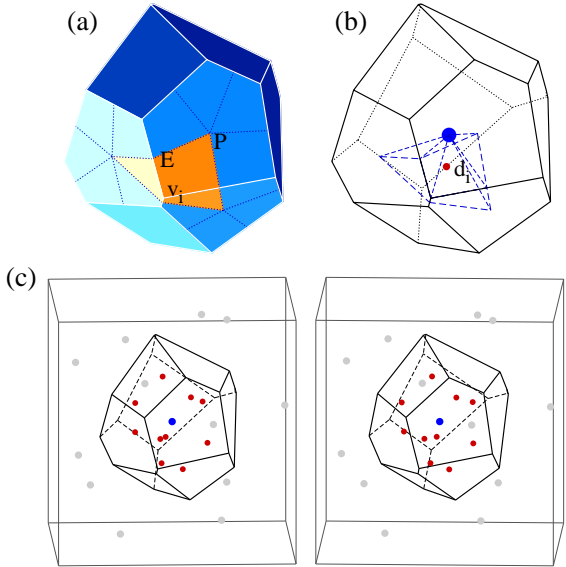}
\caption{
Top panels (a) and (b): Configuration of a subcell 
(orange region and blue-dotted lines) 
and the position of a daughter particle (red dot) 
around a parent particle (blue dot).
Bottom panel (c): Stereograph of the daughter particles 
(red dots) replacing the central
parent particle (blue dot) by our splitting method.
Black solid lines indicate the Voronoi edges and grey dots indicate the
neighbor particles around the target parent particle.
}
\label{fig:split_demo_3D}
\end{figure}

We construct the Voronoi diagram as follows.
In our scheme, the Voronoi cell for each particle is obtained.
We set a cube of size $2 l_{\rm cube}$ centered at a target 
parent particle as a first trial Voronoi cell.
The cube is then cut by a perpendicular bisector of
a line segment
which connects the central particle and each particle in the cube.
After the cube is cut by all the bisectors, we obtain the 
Voronoi cell.
Let us call this procedure the facet-cut algorithm.
Initially the cube size is set as
$l_{\rm cube} = h_{\rm parent}$.
Particles outside the initial cube but closer 
to the central particle
than $2l_{\max}$ can define 
the final shape of the Voronoi cell, where $l_{\max}$
is the maximal distance between the central particle and 
the vertices of the Voronoi cell.
If $l_{\max}$ is larger than $l_{\rm cube}/2$,
we enlarge the initial cube and repeat the facet-cut algorithm.
This algorithm scales as 
${\cal O}(N_{\rm sp} \log N_{\rm sp})$--${\cal O}(N_{\rm sp}^2)$ for the number of 
split particles $N_{\rm sp}$, with a weak dependence on the distribution 
of the particles.

There are various possibilities to determine the coordinates 
and mass of the daughter particles.
Basically, the daughter particles of a single parent 
are desired to be well separated to each other and 
to have a nearly equal mass.
Figure \ref{fig:split_demo_2D} shows an example set
of the daughter particles.
We split the Voronoi cell of the central parent particle (blue dot) 
into small subcells defined by the blue dotted lines.
A subcell corresponds to each Voronoi vertex.
A daughter is then placed at the center of mass of each subcell.
In this way, we can realize an approximately equal 
mass distribution of the daughter particles.

We extend the above procedure to three-dimensional cases. 
Figure \ref{fig:split_demo_3D} illustrates our procedure.
We begin with the configuration given in the panel (a).
We first draw a boundaries of subcells on each Voronoi 
plane by the blue dotted line between the two points $P$ and $E$,
where $P$ is the center of mass of the Voronoi plane, 
and $E$ is the middle point of the Voronoi edge.
We then define the base planes of the subcell 
corresponding to a vertex $v_i$ as the orange-colored region.
The whole shape of the subcell is defined as the polyhedron 
surrounded by the blue dotted lines in the panel (b).
The polyhedron consists of the base planes with the thick blue-dotted lines
in the panel (b), which is the same as the orange region in the panel (a),
and the target parent particle (blue dot in the panel (b)).
Next, we merge the subcells whose corresponding vertices 
are connected by the shortest edges.
We perform this merging procedure 
in order not to have daughter particles  
significantly close to each other.
A Voronoi cell in 3D typically has 20--30 vertices.
If the number of the Voronoi vertices is more than ten,
we repeatedly merge the closest subcells until the
total number is reduced to ten. At this point
we obtain a nearly uniform distribution of the daughters.
We choose the threshold number $N_{\rm th} = 10$, because we do 
not want to have a huge mass difference between the parent 
and the daughters. 
Refinement by a factor of ten is a reasonable choice. 
We have actually performed an additional simulation
in which subcells are not merged and have confirmed that
the results are not significantly
affected by the threshold number.
We finally put a daughter particle at the 
center of mass of each (merged) subcell as shown by the red dot in the panel (b).

The bottom panel in Figure \ref{fig:split_demo_3D} is
a stereogram showing the
distribution of the ten daughters (red dots) 
generated around the parent (blue dot).
We equally divide the mass of the parent particle into 
$m_{\rm daughter} = m_{\rm parent}/10$.
We give the daughters the same temperature as that of the parent.\footnote{We have a few other choices such as
assigning entropy. We choose temperature as a primary
thermodynamic quantity for simplicity.}
The velocity of the parent particle is simply
assigned to the daughters
so that the linear momentum and kinetic energy are
strictly conserved.
The daughters are given half the integration 
timestep of their parent as a temporal value
after splitting, but
the individual integration timestep is recalculated 
for the new set of particles in the next timestep.

\section{TEST SIMULATIONS}

We perform three test simulations to examine the splitting 
methods of the previous study ({\tt SPHERE})
and ours ({\tt VORO}).
We are particularly interested in 
how the new density field after splitting differs 
from the original density field represented by the
parent particles.
In the following tests, we use the SPH
code {\sc gadget}-2 \citep{Springel05}
supplemented with the parallelized algorithms of the Voronoi 
tessellation and the particle splitting.
All the control parameters are set to be 
the default values of {\sc gadget}-2.0.7.
To determine the SPH kernel size, we fix the number 
of the neighbor particles to $N_{\rm ngb} = 50 \pm 1$.
The densities of SPH particles are calculated 
in the standard manner using the cubic spline kernel.

\subsection{Random and Uniform Particle Distributions}

\begin{table}
\caption{Density deviations 
just after the particle splitting
for random particle distributions
\label{tab:TEST2}}
\begin{tabular}{@{}lrrrr}
\hline
run & 
$\rho_{\max}$ & 
$\rho_{\min}$ &
$\bar \rho$ & 
$\sigma_{\rho}$ \\
\hline \hline
{\tt UNIF-HALF-SPHERE} & 3.76 & 0.49 & 1.12 & 0.248 \\
{\tt UNIF-HALF-VORO} & 1.65 & 0.63 & 1.04 & 0.141 \\
{\tt UNIF-ALL-SPHERE}  & 4.03 & 0.72 & 1.17 & 0.223 \\
{\tt UNIF-ALL-VORO}  & 1.89 & 0.71 & 1.07 & 0.149 \\
\hline
\end{tabular}
\\ \medskip
\end{table}

We first run simple simulations of an isothermal gas without gravity.
We distribute $16^3$ SPH particles randomly in the cubic box of
$x\in [0,1)$, $y\in [0,1)$, and $z\in [0,1)$ and adopt the periodic boundary
conditions.
The total mass of the particles is 1 and thus
the mean volume-averaged density is $\bar \rho _{\rm ini} = 1$ in the code unit.
The initial particle mass is $m_{\rm p,ini} = 1/16^3$.
We set the isothermal sound speed $c_T$ such that the sound crossing time
is equal to the unit time.
Since the typical length scale 
of the density fluctuation resulting from the random
noise is comparable to the smoothing length of the particles,
we define the characteristic 
sound crossing time as $t_{\rm sc} = \bar h_{\rm ini} / c_T$, where 
$\bar h_{\rm ini}=\left(3N_{\rm ngb} m_{\rm p,ini} / 4\pi \bar \rho _{\rm ini} \right) ^{1/3}$ 
is the initial mean smoothing length.
We first run a simulation to $t=10$
without particle splitting to reduce 
the random noise by the gas
pressure.
At the time, the deviation of the density 
$\sigma _{\rho} $ is reduced down to 0.0026.
The maximal and minimal densities are found to be
$(\rho _{\max},  \rho _{\min}) = (1.08, 0.92)$.
We call the set of the runs {\tt UNIF}.

We then split half of the particles in $x\in [0,0.5)$ for run {\tt HALF}
and the all particles for run {\tt ALL}.
Each set is performed with the two splitting methods of {\tt SPHERE}
and {\tt VORO}.
After particle splitting, the root-mean-square 
of the density $\sigma _{\rho}$ suddenly increases;
the maximal density increases and also the minimal density decreases.
Note that the particle-averaged density fluctuates
around unity after perturbed by the particle splitting.
The density deviation eventually approaches to zero and the minimal 
and maximal densities 
approach to the mean density because of the gas pressure.
We summarize the results in Table \ref{tab:TEST2}. 
The second to fifth columns show the maximal, minimal, average, 
and root-mean-square of the
density just after the particle splitting, respectively. 

For the run {\tt HALF}, the density deviation from 
the initial value is smaller with {\tt VORO} than with {\tt SPHERE}.
The maximal and minimal densities just after the splitting change
from 
$(\rho _{\max},  \rho _{\min}) = (3.76, 0.49)$ to $(1.65, 0.63)$, 
with the latter being closer to 1.
The density deviation is also reduced from 0.25 to 0.14.
The average density is larger than the original density 
by 12 \% for {\tt SPHERE}, while 4 \% for {\tt VORO}.
In the region $x\in (0.1, 0.4)$ populated with the fine particles,
the density is overestimated at most by a factor of 4 and 
underestimated by a factor of 0.3 for {\tt SPHERE},
whereas the density is overestimated by a factor of 0.6 
and underestimated by a factor of 0.4 for {\tt VORO}.
At the boundary of the split and un-split particles, 
the density of the coarse particles is overestimated
by a factor of 2.5 and underestimated by a factor 
of 0.5 for {\tt SPHERE}, 
while the density deviates by a factor of 0.3 for {\tt VORO}.
At the time $t=20$, when we stop the simulations,
the density difference still remains at the boundary.
The maximal and minimal densities are 0.3 and 0.2 
for {\tt SPHERE} and {\tt VORO}, respectively.
Because the particles 
with different masses are mixed near the boundary,
the local density does not quickly converge to the mean
value.
In practical gas collapse simulations, however, the
densest cloud core shrinks more rapidly than the 
boundaries of fine and coarse particles.
Thus the remaining density perturbation on the boundaries 
would have little effect on the central regions.

For the run {\tt ALL}, the original density field is well 
preserved by our splitting method.
The maximal density is 1.89 for {\tt VORO}, which is 
by a factor of two better than {\tt SPHERE}.
Although the density is slightly more underestimated 
for {\tt VORO}.
the overestimation of the average density is improved from 17 \% to
only 7 \%, and the root-mean-square of the density is reduced 
from 0.22 to 0.15.
As we have seen in Section \ref{sec:intro}, the density is 
overestimated for a particle that is placed 
significantly close to one of the neighbor particles.
This happens with the {\tt SPHERE} configuration because it involves
randomization of the orientation of the daughter particle distribution.
With {\tt VORO}, such an unfortunate case with significant density 
overestimation does not occur because the coordinates of the daughters
are uniquely determined with
the separation of any two daughters 
being a fraction of the distance of their parents.
We continue the run and find that 
another $4.6$ sound-crossing time is required 
for the density deviation to recover to the initial value 0.026
for both {\tt SPHERE} and {\tt VORO}.

\subsection{Particle Distribution with Density Gradient}
\label{subsec:grad}

\begin{figure}
\includegraphics[width=8cm]{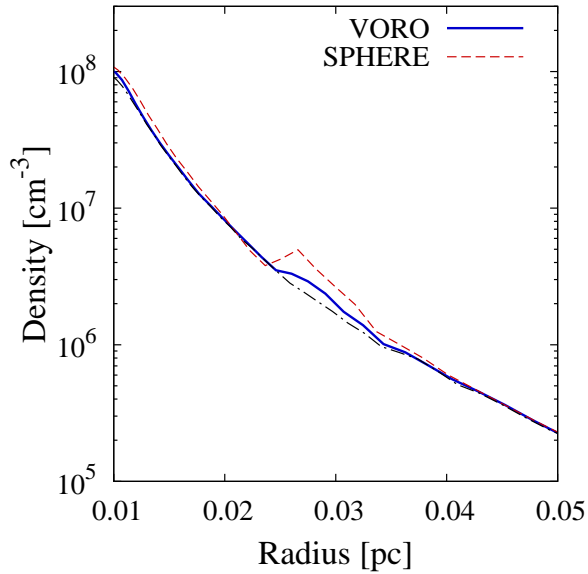}
\caption{
Density profile as a function of the radius for our test {\tt GRAD}.
The black dot-dashed curve shows the density profile 
before the particle splitting.
The solid and dashed lines show the density profiles
just after the particle splitting
with the methods of the previous (red dashed) and this (blue solid)
works, respectively.
}
\label{fig:grad}
\end{figure}

It is important to test our method in the case 
with a steep density gradient.
If the length scale of a density fluctuation 
$l_{\rho} = \rho / |d\rho/dr |$ is smaller than
$h_{\rm parent} \sim (N_{\rm ngb} m_{\rm p}/\rho)^{1/3}$, 
the structure with length scales $\lesssim h_{\rm parent}$ 
would be smoothed out with the previous splitting method.
On the other hand, with our method, 
the length scale over which the daughter particles 
are distributed is comparable
to the inter-particle distance
$\lambda _{\rm p} \sim (m_{\rm p}/\rho) ^{1/3}$ before split.
Thus, the length over which the density profile
is affected is expected to be much smaller.
To illustrate this, let us 
consider a spherical cloud with a large density 
gradient $\rho \propto r^{-4}$, 
corresponding to small $l_{\rho}$.
Such structure can be seen on the edge of 
the disk in rotating isothermal clouds (Section \ref{subsec:bb79}),
and also in a volume within a primordial gas cloud where 
the transition lines of HD molecules becomes
optically thick (see Section \ref{sec:mh}).
We call this test {\tt GRAD}.

We realize the desired density profile 
with $\rho \propto r^{-4}$ by 
perturbing particles that are homogeneously distributed 
initially.
The black dot-dashed curve in Figure \ref{fig:grad} 
shows the generated density profile.
The sound speed of the cloud is $c_s = 1 \ {\rm km \ s^{-1}}$, 
corresponding to $T=100$ K, actually being 
close to the primordial gas cloud temperature.
Then, we split the particles with densities 
$\nH > 3\E{6} \ \percc$.
At the density, the length scale of the density 
variation is $l_{\rho} = r/4 = 0.006$ pc.
The red dashed curves and blue solid curves in Figure \ref{fig:grad} show the 
density profiles just after the particle splitting of the previous and our works.
The density profile is flattened at $r=0.03$ pc over the length scale
$\sim h_{\rm parent} =  0.01$ pc for {\tt SPHERE},
while we find only a small bump with $\sim \lambda _{\rm p} \sim 0.004$ pc
with our method.
The daughter particles located within the inter-particle distance 
of their parent can preserve the density structure before splitting.
Note that we consider a spherical cloud in this test.
For a cloud with anisotropic density structures, 
distributing the daughters isotropically results 
in smoothing the original density structure.
The artificial smoothing is mitigated with our method,
as we will see in the case of a primordial cloud in Section \ref{sec:mh}.

\subsection{Bodenheimer test}
\label{subsec:bb79}

\begin{figure}
\includegraphics[width=8cm]{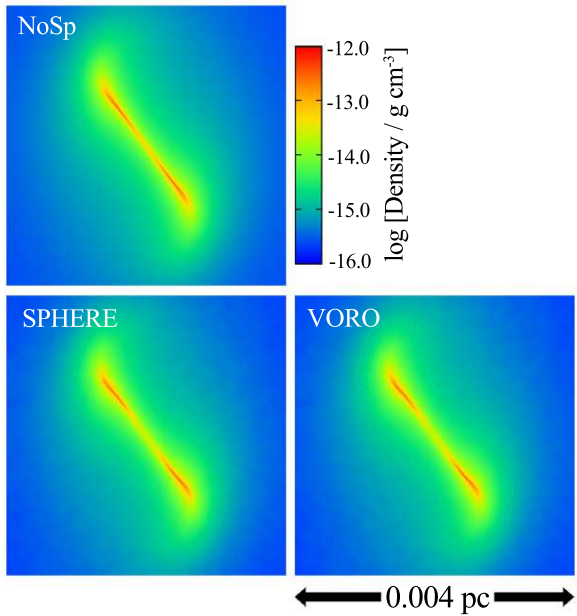}
\caption{
Density slice on the equatorial plane for the Bodenheimer test simulations 
without particle splitting ({\tt NoSp}: top)
and with the particle splitting of the previous work  ({\tt SPHERE}: bottom left) 
and the present work ({\tt VORO}: bottom right) 
when the maximal density is $\sim 5\E{12} \ \percc$. 
}
\label{fig:BB79_snapshots}
\end{figure}

\begin{figure}
\includegraphics[width=8cm]{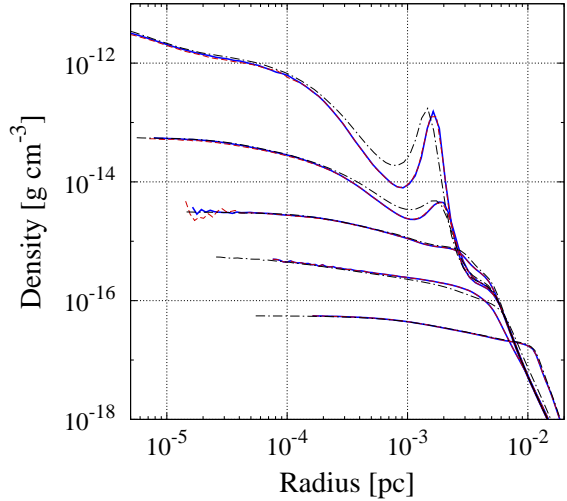}
\caption{
Radial profile of the density for Bodenheimer test simulations without the particle splitting ({\tt NoSp}: black dot-dashed) and
with the particle splitting of the previous work ({\tt SPHERE}: red dashed) and the present work ({\tt VORO}: blue solid).
Note that the red and blue curves closely 
lie on each other in the plot.
}
\label{fig:BB79_rn}
\end{figure}

\begin{table}
\caption{Density deviations for {\tt BB79} tests
\label{tab:BB79_results}}
\begin{tabular}{@{}lrrrrr}
\hline
run & 
$N_{{\rm p, ini}}$ &
$N_{{\rm p, fin}}$ &
$\frac{\rho_{\max} }{ \bar \rho }$ & 
$\frac{\rho_{\min} }{ \bar \rho }$  &
$\frac{\sigma_{\rho} }{ \bar \rho } $ \\
\hline \hline
{\tt BB79-NoSp} & 113,098,912 &  &  &  & \\
{\tt BB79-SPHERE}  & 4,188,896 & 15,925,664 & 5.54 & 0.41 & 0.33 \\
{\tt BB79-VORO} & 4,188,896 & 12,805,187 & 2.22 & 0.62 & 0.15 \\
\hline
\end{tabular}
\\ \medskip
\end{table}

Next, we perform a well-known test introduced by \citet{Boss79} ({\tt BB79}).
We follow gravitational collapse of an isothermal cloud with and without the particle splitting.
The cloud mass is $M_{\rm ini} = 1 \ \Msun$ and its 
radius is $R_{\rm ini} = 0.02$ pc initially.
The isothermal sound speed of the gas is $c_T = 0.17 \ {\rm km \ s^{-1}}$.
The ratio of the thermal energy to the gravitational 
energy is $\alpha _{\rm ini} = 
5c_T^2 R_{\rm ini} / 2GM_{\rm ini} = 0.32$.
We set the initial angular velocity 
$\Omega _{\rm ini} = 7.2\E{-13} \ {\rm s^{-1}}$
and then the ratio of the rotational energy to the gravitational energy is
$\beta _{\rm ini} = \Omega _{\rm ini}^2 R_{\rm ini}^3 /3GM_{\rm ini} = 0.28$.
The amplitude of $m=2$ perturbation is $A_{m=2, {\rm ini}} = 0.1$.
The particles are distributed initially on a regular grid, 
and the particle mass is modulated as
$m_{{\rm p, ini}} = \bar m_{{\rm p, ini}} (1+A_{m=2, {\rm ini}} \cos 2 \phi / 2)$,
where $\bar m_{{\rm p, ini}}$ is the initial average particle mass, and $\phi$
is the azimuthal angle of the particles.

The semi-analytic approaches by \citet{Tsuribe99b} show that,
in the cloud with $(\alpha, \beta) = (0.32, 0.28)$, 
the non-axisymmetric structure of the gas evolves, and
the fragmentation property depends on the initial amplitude of 
the perturbation.
\citet{Tsuribe99a} perform the three-dimensional simulations 
and conclude that such cloud fragments if
the initial amplitude is larger than 0.15.
In the {\tt BB79} test, we run simulations with the minimal 
mass resolutions $\Rcr =1$, 10, and 100.
For $\Rcr =1$ and 10, artificial fragmentation occurs 
for the lack of the resolution
as suggested by \citet{Truelove97}.
We here show the results for $\Rcr =100$.

We perform three runs and compare them.
In the run {\tt NoSp}, we distribute a very large
number of particles ($N_{\rm p} \sim 10^8$)
to satisfy the Jeans criterion ${\cal R} \equiv M_{\rm Jeans}/M_{\rm min} \gtrsim 100$ 
until the gas density reaches $5\E{-12} \ {\rm g \ cm^{-3}}$,
where the gas would become optically thick.
In the runs {\tt SPHERE} and {\tt VORO}, we initially distribute 
$N_{\rm p,ini} \sim 4\E{6}$ particles.
When particles meet the splitting criteria, i.e., violating
the Jeans condition, they are split one by one in the course of the 
simulations.

Figure \ref{fig:BB79_snapshots} 
shows the density slice of the clouds for
{\tt NoSp} (top), {\tt SPHERE} (bottom left), and {\tt VORO} (bottom right)
at the maximal density $\rho _{\rm cen} = 5\E{-12} \ {\rm g \ cm^{-3}}$.
The stretched thin filament does not fragment 
in all the three runs up to the last output time as reported by
\citet{Tsuribe99b}.
Figure \ref{fig:BB79_rn} shows the evolution of the 
density profiles 
as a function of the distance from the densest part of one 
end of the filament.
The overall structure of the filament remains essentially
the same among the three runs.
We conclude that the resolution $\Rcr=100$ is 
sufficient when we consider collapse and fragmentation
of the cloud.
The number of particles is $N_{\rm p,fin} \sim 10^7$ for 
{\tt SPHERE} and {\tt VORO} when we terminate the simulations
at $\rho_{\rm cen}= 5\E{-12} \ {\rm g \ cm^{-3}}$, which is by an order of magnitude
less than {\tt NoSp}.
This means that the computational cost is well reduced by the splitting technique
although the final snapshots are similar to the one for {\tt NoSp}.

Let us compare the two runs {\tt SPHERE} and {\tt VORO}
in detail.
We follow the density deviations of the 1000 particles that are 
split first.
The results are summarized in Table \ref{tab:BB79_results}.
The particles are split for the first time 
when $\rho = 3\E{-15} \ {\rm g \ cm^{-3}}$.
Just before the splitting, the maximal and minimal densities 
relative to the mean density of the 1000 particles are
1.08 and 0.85, respectively, 
and the deviation of the density is 0.037.
Just after the 1000 particles are split,
the density deviation marks the maximum.
In the {\tt SPHERE} and {\tt VORO} runs, the maximal and
minimal densities just after the particle splitting are
$(\rho _{\max}/\bar \rho, \rho _{\min}/\bar \rho ) = (5.54, 0.41)$ and $(2.22, 0.62)$, respectively.
The maximal deviations of the density from the original value 1 are reduced
more than by a factor of two by our splitting method.
The density deviation relative to the mean density just after the splitting
also decreases from 0.33 ({\tt SPHERE}) to 0.15 ({\tt VORO}),
approaching to the initial value of 0.037.
The result shows that,
also in the practical collapse simulation, 
the daughter particles distributed in a Voronoi cell
of each split particle are not close to other neighbor particles,
and their density is not significantly overestimated,
while this happens with the conventional splitting method.

In this {\tt BB79} test, the density structure is not 
very different between {\tt SPHERE} and {\tt VORO},
showing an interesting 
contrast to our test {\tt GRAD} that is more relevant 
to a primordial warm gas (see Section \ref{subsec:grad}).
In the cold gas with small $c_T$ in {\tt BB79}, 
the smoothing length $h_{\rm parent} \sim c_T \Rcr ^{-1/3} (G\rho )^{-1/2}$ 
is smaller than the length scale $l_{\rho}$ 
all over the region if the critical condition is set as $\Rcr =100$.
On the iso-density surface of $\rho = 5\E{-13} \ {\rm g \ cm^{-3}}$, where
the second particle splitting occurs,
$l_{\rho}$ is smallest, $4\E{-6} \ {\rm pc}$, at the edge 
of the iso-density filament.
The smoothing length in the region is 
$h_{\rm parent} = 3\E{-6} \ {\rm pc}$.
SPH particles which satisfy the Jeans criterion $\Rcr =100$ can resolve
the density structure when the gas is sufficiently 
cold ($c_T\sim 0.1 \ {\rm km \ s^{-1}}$).
This is, however, not in the case of the primordial warm gas 
as we revisit with detailed discussion later in Section \ref{sec:mh}.

\section{Collapse simulations of primordial gas cloud}
\label{sec:mh}

\begin{figure}
\includegraphics[width=8cm]{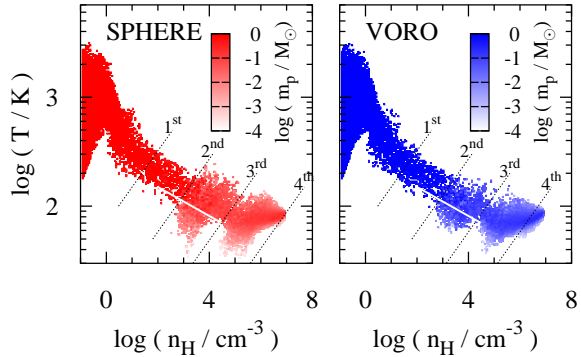}
\caption{
Distribution of the particles 
on the thermodynamic phase plane
in the simulations of primordial gas cloud formaton, 
when the central density $n_{\rm H, cen} = 10^7 \ \percc$.
The color indicates the mass-weighted number 
of the particles.
Left and right panel show the result  
with the particle splitting of {\tt SPHERE} and {\tt VORO}, respectively.
The dotted lines attached to the orders `$n$th' indicate 
the density and temperature where the particles are split
for the $n$th time.
The SPH particles in the region below the white diagonal line 
have temperatures that are underestimated 
due to the particle spitting.
}
\label{fig:MH_nT}
\end{figure}

\begin{figure}
\includegraphics[width=8cm]{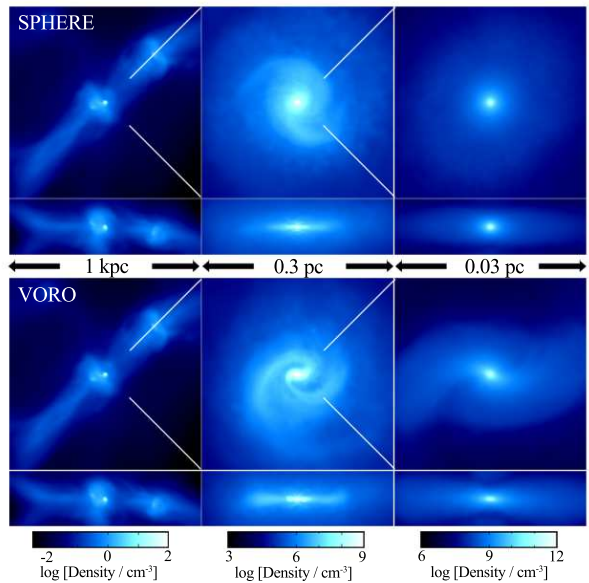}
\caption{
Projected density distribution of the primordial clouds from the face-on and 
edge-on views of the disk
methods of {\tt SPHERE} (top) and {\tt VORO} (bottom) 
when the central density is $10^{12} \ \percc$.}
\label{fig:MH_snapshots}
\end{figure}

\begin{figure}
\includegraphics[width=8cm]{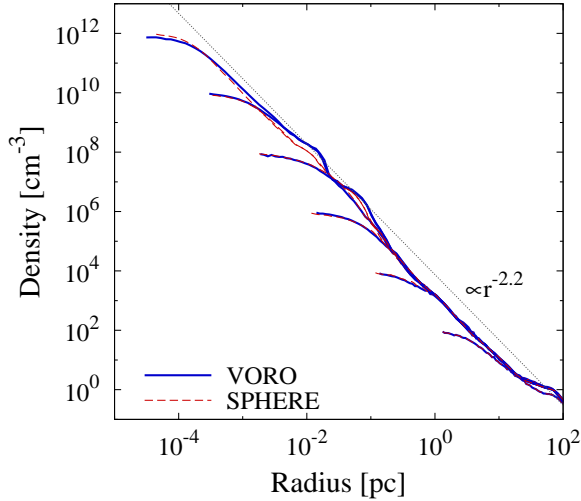}
\caption{
Radial density profile of the primordial cloud 
with the particle splitting methods
{\tt SPHERE} (red dashed) and {\tt VORO} (blue solid).
We also plot the grey dotted line $\propto r^{-2.2}$ which a primordial gas cloud tends to be along.}
\label{fig:MH_rn}
\end{figure}

Finally, we apply our splitting method to a more
realistic calculation of the collapse 
of a primordial minihalo.
We isolate the spherical region with length $\sim 1$ kpc 
centered at one of the minihalos
formed at the redshift $z=15$ in the cosmological 
simulation of \citet{Hirano14}.
The mass of the halo is $M_h = 4.8\E{5} \ \Msun$.
The halo is categorized to a subclass of the primordial minihalos 
which hosts less massive
stars ($10$--$100 \ \Msun$) because of the efficient cooling by HD molecules.
The chemical reaction rates and radiative cooling rates are taken 
from \citet{Chiaki15}
for a primordial composition.
We reduce the cooling rates of emission lines by the Sobolev length method in
the optically thick regime as in \citet{Hirano13}.
The splitting methods of the privious work ({\tt SPHERE}) 
and this work ({\tt VORO}) are employed for 
the gas particles.
The number of neighbor particles is restricted to be 
$N_{\rm ngb} = 64 \pm 8$, and the criterion for the particle splitting 
is $\Rcr =1000$ as set by \citet{Hirano14}.
The central particles are already once split 
by {\tt SPHERE} splitting when the zoomed initial conditions are
generated. The initial particle mass is
$m_{\rm SPH}^{(1)} = 5/13 \ \Msun$.
The number of gas particles is initially 169,511.
When the central density reaches $\nH = 10^{12} \ \percc$, 
the particles have been split at most five times 
and the number of split particles increases up to
314,867
for {\tt SPHERE} and
318,148
for {\tt VORO}.

Figure \ref{fig:MH_nT} shows that distribution of the
gas particles in the density-temperature plane. 
We use the snapshot when the central density $n_{\rm H,cen}=10^7 \ \percc$.
Note that HD cooling operates at the densities $\nH \gtrsim 10^3 \percc $,
where a disk structure is formed with noticeable spiral arms.
The particles are split at the densities and temperatures indicated by the
dotted lines in Figure \ref{fig:MH_nT}.
The index `$n$th' attached to the dotted lines indicates the criterion for the
$n$th particle splitting: $M_{\rm Jeans} = \Rcr N_{\rm ngb} (m_{\rm SPH}^{(1)} / 13^{n-1})$ and
$\Rcr N_{\rm ngb} (m_{\rm SPH}^{(1)} / 10^{n-1})$ for {\tt SPHERE} and {\tt VORO}, respectively.
We see that the temperature significantly fluctuates 
along the adiabatic line $T\propto \nH ^{2/3}$
from the splitting points for {\tt SPHERE}.
Since the gas cooling and heating do not work promptly within a
short timestep,
the temperatures of daughters deviate along the adiabatic line when their densities
are perturbed from the parent's density one timestep after splitting.
Thus the temperature deviation almost directly reflects 
the density perturbation induced by the particle splitting.
The white lines in Figure \ref{fig:MH_nT}  
indicate the expected trajectories,
i.e., the gas particles having the lowest temperature
at each density would evolve roughly on the white line
without the particle splitting. 
The figure shows that our splitting method slightly improves the temperature 
deviation from the splitting point.
The distributions of particles on the density-temperature planes are not
significantly different for two splitting methods {\tt SPHERE} and {\tt VORO}.

We find that substantial differences are caused
in the global structure by the two different splitting methods.
Figure \ref{fig:MH_snapshots} shows the snapshots of the gas cloud 
at the central density 
$n_{\rm H, cen}=10^{12} \ \percc$.
Comparing the middle panels, we find that the spiral arms
have more developed 
in {\tt VORO} than in {\tt SPHERE}.
This is related to the associated smoothing and
isotropization after split, as we discussed in Section \ref{subsec:grad}.
At densities $\nH\sim 10^6$--$10^7 \ \percc$, 
where the emission lines from HD molecules becomes
optically thick, $l_{\rho}$ is reduced to $\sim 0.01$ pc 
because of the rapidly increasing pressure.
The length at which the daughter particles are distributed is 
$\sim h_{\rm parent} \sim (N_{\rm ngb} m_{\rm p}/ \rho )^{1/3} \simeq 0.01$ pc 
in run {\tt SPHERE} and
$\sim \lambda _{\rho} \sim (m_{\rm p}/\rho)^{1/3} \simeq 0.004$ pc 
in run {\tt VORO}.
Even for the same Jeans criterion, 
our splitting method allows to make the resolution of daughter particles 
effectively higher than in {\tt SPHERE} 
by a factor of $\sim h_{\rm parent} / \lambda _{\rm p} \sim N_{\rm ngb}^{1/3}$.

Figure \ref{fig:MH_rn} shows the radial density profile
and its evolution.
Our splitting method maintain the steep density profile
at the densities $\nH \sim 10^6 \ \percc$ where the 
emission lines of HD molecules are 
optically thick, and
at $\sim 10^8 \ \percc$ where the gas heating by H$_2$ molecular 
formation via the three-body reactions is dominant.
The right panels of Figure \ref{fig:MH_snapshots} show that 
the bar-like structure has developed
in {\tt VORO} while the central region appears smooth
and spherical in {\tt SPHERE}.
Clearly, details of splitting method affect
the evolution of the global structure
such as spiral arms and also the fragmentation
of the cloud core.

We suspect from the tests {\tt GRAD} and {\tt BB79} that
the anisotropic density structure in the primordial 
collapsing gas is smoothed by the 
isotropic distribution of daughter particles.
In a warm gas cloud with large $c_T$ such as a primordial gas cloud,
the length $l_{\rho}$ can not be resolved
by $h_{\rm parent} \propto c_T \Rcr ^{-1/3} (G\rho)^{-1/2}$ even though
the Jeans criterion with $\Rcr =1000$ is satisfied.
To avoid the isotropic diffusion of density structures in a warm gas 
where $c_T\gtrsim 1 \ {\rm km \ s^{-1}}$, 
we should impose a severer Jeans criterion 
with larger resolution threshold $\Rcr$ 
such that $l_{\rho}$ can be resolved
by the coarse particles with the conventional splitting method (see Section \ref{sec:discussion}).
Our method overcomes this problem by distributing the fine particles 
within an inter-particle
distance from the parent, which shortens the effective resolution 
of the gas particles.
Therefore, it is not necessary to have 
severer criteria for splitting than the Jeans criterion.

\section{DISCUSSION}
\label{sec:discussion}

\begin{figure}
\includegraphics[width=8cm]{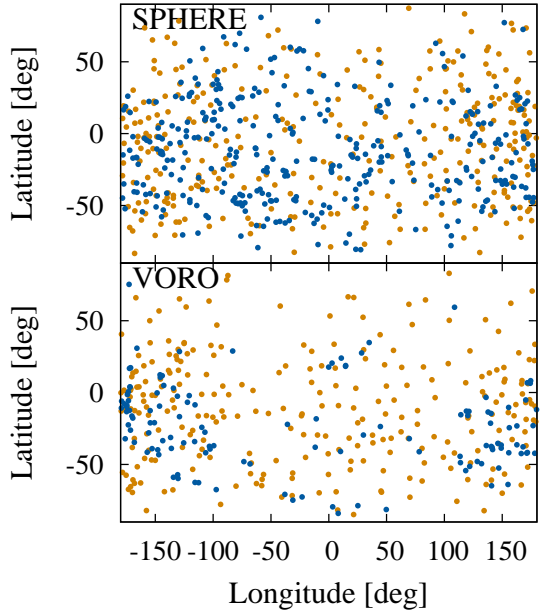}
\caption{
Distribution of the SPH particles above (orange) and below (blue) the white line in Figure \ref{fig:MH_nT}
on the iso-density sphere with $\nH = 10^3 \ \percc$ for the splitting methods of the previous (top) and this work (bottom).}
\label{fig:MH_ta}
\end{figure}

\begin{figure}
\includegraphics[width=8cm]{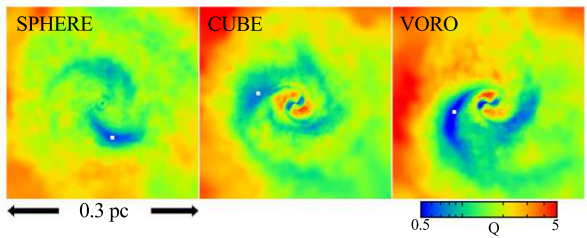}
\caption{
Distribution of the Toomre $Q$ parameter in the disk
from the face-on view for runs {\tt SPHERE} (left), {\tt CUBE} (middle) and {\tt VORO} (right)
in the same region as the middle panels of Figure \ref{fig:MH_snapshots}.
The $Q$ parameter is smallest at the white squares in the spiral arms.
}
\label{fig:MH_Q}
\end{figure}

We have developed a novel method for particle splitting 
in SPH simulations based on the Voronoi diagram ({\tt VORO}).
Our series of tests show that the new method preserves
well the original density structure before splitting,
while achieving an effectively high resolution.
The density perturbations induced by the splitting
are reduced by a factor of two compared with the conventional 
splitting method ({\tt SPHERE}).
We have found that similar improvements are achieved by 
another particle splitting method of MES06 ({\tt CUBE} hereafter), 
where daughters are distributed within a distance $h_{\rm parent}/8$ 
($\simeq \lambda _{\rm p}/4$) from their parent. 
In the test {\tt GRAD}, the density profile is similar to {\tt VORO}.
Only the slight bump appears at the splitting point over the 
length scale $\sim \lambda _{\rm p}$.
In the test {\tt BB79}, the morphology and the density profile is similar to
runs {\tt SPHERE} and {\tt VORO}.
The density deviation just after the splitting is better than {\tt VORO}.
The maximal and minimal densities relative to the average density
of 1000 particles first split are 1.91 and 0.68, respectively.
The root-mean-square of the density is 1.03.

In the case of a contracting primordial gas cloud,
where the gas no longer evolves isothermally,
the temperature deviates along with the density perturbation
generated by the particle splitting as Figure \ref{fig:MH_nT} shows.
Some particles are overcooled and might trigger the artificial
fragmentation by their small pressure, but this would not be in the case.
Figure \ref{fig:MH_ta} shows the spatial distribution of the SPH particles
on the iso-density surface with $\nH = 10^3 \ \percc$.
We divide the gas particles into hot (orange) and cool (blue) components 
respectively above and under the white line in Figure \ref{fig:MH_nT}.
The blue dots indicate the overcooled gas.
If the overcooled gas particles cluster, they might trigger the fragmentation
by their small pressure.
However, the cool gas particles apparently scatter 
in all directions for {\tt SPHERE} (upper panel).
For {\tt VORO}, although the cool component appears to cluster 
at longitude $180^{\circ}$, where a spiral arm lies,
the hot gas particles are also in the arm region and
would prevent the collapse of the cool component.
The local perturbations of temperature and density triggered
by the particle splitting would not affect the gas fragmentation
irrespective of the splitting methods.

On the other hand,
the global density structure such as spiral arms
and bars can affect the gravitational stability of the gas.
As seen in Figure \ref{fig:MH_snapshots}, 
the high-density region remains 
strongly elongated in {\tt VORO} run.
Although the long-time evolution of the spiral arms is not followed 
in our simulations,
we can derive a reasonable estimate on whether or not the disk is stable
by calculating the Toomre $Q$ parameter $Q=2\Omega c_T/\pi G \Sigma$, where
$\Sigma $ is the column density in each part on the disk.
Figure \ref{fig:MH_Q} shows the distribution of $Q$ parameter. 
Clearly, the value of $Q$ is small along the spiral arms 
and is the smallest 
at the point indicated by the white square in each panel:
$Q_{\min} = 0.59$ for {\tt SPHERE} and 0.41 for {\tt VORO}.
We have also performed the collapse simulation with the splitting
method {\tt CUBE}.
The distribution of $Q$ parameter is shown in the middle panel of Figure \ref{fig:MH_Q}.
The elongation of the central core is slightly smaller than our method {\tt VORO}.
The inner structure of the spiral arms is resolved as well as {\tt VORO}, but
the density structure appears slightly more diffuse. This is likely 
because of the nature of the method of
{\tt CUBE}, where the daughter particles are distributed on the vertices of the cube
around a parent, and the original density structure formed by parent particles are
slightly smoothed in the directions both perpendicular and parallel to the disk.
The column density in the spiral arms for the run {\tt CUBE} is less than {\tt VORO},
and the resulting minimal $Q$ parameter is 0.58.
The linear analysis of the ring-mode instabilities by \citet{Takahashi15} 
suggests that spiral arms with $Q \lesssim 1/\pi$ is
dynamically unstable and likely produce fragments.
We can say that the spiral arms are marginally
stable against fragmentation in the both runs at the time of the snapshot.
Still, the difference in the global structure of the protostellar
disk affects the evolution of the central protostar in a complicated
manner \citep[see e.g.][]{Greif12, Vorobyov13}. 
It is important to fully resolve the disk structure over a long time.

Our test simulations suggest
that the mass resolution with  {\tt VORO} is effectively 
higher than with {\tt SPHERE}
by a factor of $\sim h_{\rm parent} / \lambda _{\rm p} \sim N_{\rm ngb}^{1/3}$.
We find that small-scale anisotropic structures are smoothed 
out when the particles are split
in the region with steep density gradient with insufficient resolution 
with {\tt SPHERE} method.
To avoid the artificial mixing,
we suggest to impose an additional criterion than the Jeans criterion
for splitting with the conventional splitting 
method with spherical distribution
of daughters:
the length scale of the density fluctuation $l_{\rho}$ 
should be sufficiently resolved by the local SPH smoothing length. 
To confirm this, we have performed another simulation 
with {\tt SPHERE} splitting with a substantially
higher resolution of $\Rcr = 4000$.
The high-resolution run indeed shows an elongated cloud 
and apparent spiral arms, similarly to those found in the {\tt VORO} run.
We emphasize that our splitting method performs
only with the modest Jeans criterion,
because the distribution of daughters traces  the original anisotropic 
density structure before splitting.
The method saves computational cost by effectively reducing the
number of particles necessary to resolve the cloud structures.

It is actually costly to construct the Voronoi diagrams 
with the current algorithm.
In the simulations with $\Rcr =1000$, the computational time 
until the density of the cloud center reaches $n_{\rm H,cen}=1\E{12} \ \percc$
is 48,315 ({\tt SPHERE}) and 81,320 ({\tt VORO}) seconds using 128 cores.
The Voronoi tessellation accounts for about a half of the total 
computational time.
However, in the simulation of {\tt SPHERE} with $\Rcr =4000$,
where the bar-like structure is seen also with the conventional splitting method,
the computational time is 323,309 seconds, exceeding the time 
for {\tt VORO} with $\Rcr =1000$.
Clearly, the formation of small-scale structrue such as 
the bar-like structure is accurately followed with
 smaller computational time with
our splitting method.
We can further save the computational 
cost for generation of the Voronoi tessellation 
by introducing the incremental construction
algorithm scaling as ${\cal O}(N\log N)$ \citep{Sugihara94} .
In a forthcoming paper, we present an efficient scheme 
to parallelize the algorithm that scales as
${\cal O}(N)$ (Chiaki, Yoshikawa \& Yoshida in prep.).

We note that there have been some simplifications 
adopted in our method.
First, we place the daughter particles at the center 
of the mass of the subcells as shown
in Figure \ref{fig:split_demo_3D}.
Another promising choice might be to place daughters
such that the original Voronoi planes between 
the parent and the neighboring particles are not affected
even after the splitting.
Second, we still follow the traditional SPH scheme to solve the 
equation of motion.
\citet{Saitoh13} point out that the traditional SPH scheme
causes asymmetric errors of density and pressure estimations 
at boundaries between the two phases with different densities.
The density and pressure errors obviously affect the fluid motion.
We find that, in the case of our particle splitting, 
the pressure perturbation at the boundary between the 
fine and coarse regions nearly cancel out
and thus spurious force is not induced.
Third, the distribution of the parent particles 
has intrinsic noise due to its discrete nature.
Then unphysical anisotropic shapes of the Voronoi cells
might generate and even promote density perturbations.
We could regulate the shapes of the Voronoi cells 
by Lloyd's algorithm \citep{Lloyd82}, 
which is expected to yield accurate density 
estimates.
Finally, we adopt the cubic spline kernel 
to estimate the density as in the standard SPH.
The choice of other kernel functions and the number
of the neighbor particles would
affect the accuracy of the density reconstruction.
Overall, we have shown that our splitting method based on 
the Voronoi diagram improves the 
preservation of the density 
and the morphology of gravitationally contracting
clouds even with the minimal implementation,
but further studies are certainly warranted
to optimize the choice and the implementation
of each part of the splitting scheme.

The splitting method 
can be applied not only to simulations of star forming clouds 
but also to other astrophysical problems.
For example, in the downstream of the cooling shocks, the density 
rapidly increases by three orders
of magnitude \citep{Chiaki13}.
The density structure is highly anisotropic 
around the shock front by hydrodynamic
instabilities. 
The dense shocked region can be fully resolved 
by our splitting method while maintaining the fine structure. 
We can also perform de-refinement 
of the particles as the converse operation of the 
particle splitting.
The particles to be merged can be uniquely identified once 
we define the Voronoi diagram.
An example implemented in a moving-mesh code
is found in {\sc arepo} \citep{Springel10}.
The de-refinement can also be applied to the problem 
of an interstellar shock, where the density of 
fluid elements decreases after passing through the shocked region.
In such cases, de-refinement of particles behind 
the shock will significantly save the number of particles
and the overall computational time.
We continue developing the de-refinement techniques
based on the Voronoi diagrams.
Our method of particle spitting presented here and de-refinement technique
can be applied to particle-based simulations of various problems
as elaborated schemes both to preserve complex 
density structures and to save the
computational cost.

\section*{acknowledgments}

We thank Shingo Hirano for giving us the simulation data, and
Takayuki Saitoh, Kohji Yoshikawa, and Sanemichi Takahashi for the fruitful discussions.
GC is supported by Research Fellowships of the Japan
Society for the Promotion of Science (JSPS) for Young Scientists.
NY acknowledges the financial supports from JST CREST
and from JSPS Grant-in-Aid for Scientific Research (25287050).
The numerical simulations are carried out on Cray XC30 at Center for Computational Astrophysics, National 
Astronomical Observatory of Japan and 
on COMA at Center for Computational Sciences in University of Tsukuba.


\end{document}